\documentclass[sigconf]{acmart}

\usepackage{multirow}
\usepackage[inline]{enumitem}
\usepackage{balance}

\AtBeginDocument{%
  }

\setcopyright{rightsretained}
\copyrightyear{2024}
\acmYear{2024}
\acmDOI{XXXXXXX.XXXXXXX}

\acmBooktitle{Communications of the ACM}

\begin{abstract}
As many of us in the information retrieval (IR) research community know and appreciate, search is far from being a solved problem. Millions of people struggle with tasks on search engines every day. Often, their struggles relate to the intrinsic complexity of their task and the failure of search systems to fully understand the task and serve relevant results \cite{white2018opportunities}. The task motivates the search, creating the gap/problematic situation that searchers attempt to bridge/resolve and drives search behavior as they work through different task facets. Complex search tasks require more than support for rudimentary fact finding or re-finding. Research on methods to support complex tasks includes work on generating query and website suggestions \cite{hassan2014supporting,white2007studying}, personalizing and contextualizing search \cite{bennett2012modeling}, and developing new search experiences, including those that span time and space \cite{white2019multi,agichtein2012search}. The recent emergence of generative artificial intelligence (AI) and the arrival of assistive \emph{agents}, based on this technology, has the potential to offer further assistance to searchers, especially those engaged in complex tasks \cite{10.1145/3539618.3593069,10.1145/3583780.3615317}. There are profound implications from these advances for the design of intelligent systems and for the future of search itself. This article, based on a keynote by the author at the 2023 ACM SIGIR Conference, explores these issues and how AI agents are advancing the frontier of search system capabilities, with a special focus on information interaction and complex task completion.
\end{abstract}

\begin{document}

\title{Advancing the Search Frontier with AI Agents}

\author{Ryen W. White}
\affiliation{%
  \institution{Microsoft Research}
  \city{Redmond}
  \state{WA}
  \country{USA}
  \postcode{98052}
}
\email{ryenw@microsoft.com}
\renewcommand{\shortauthors}{White}

\maketitle

\section{Taking Search to Task}

Tasks are a critical part of people's daily lives. The market for dedicated task applications that help people with their ``to do'' tasks is likely to grow significantly (effectively triple in size) over the next few years.\footnote{\url{https://www.verifiedmarketresearch.com/product/task-management-software-market/}} There are many examples of such applications that can help both individuals (e.g., Microsoft To Do, Google Tasks, Todoist) and teams (e.g., Asana, Trello, Monday.com) tackle their tasks more effectively. Over time, these systems will increasingly integrate AI to better help their users capture, manage, and complete their tasks \cite{white2022intelligent}. In information access scenarios such as search, tasks play an important role in motivating searching via gaps in knowledge and problematic situations \cite{belkin1980anomalous,dervin1998sense}. AI can be central in these search scenarios too, especially in assisting with complex search tasks.

\subsection{Tasks in Search}

Tasks drive the search process. The IR and information science communities have long studied tasks in search \cite{shah2023taking} and many information seeking models consider the role of task directly \cite{belkin1980anomalous,dervin1998sense}. Prior research has explored the different stages of task execution (e.g., pre-focus, focus formation, post-focus) \cite{vakkari2001theory}, task levels \cite{savolainen2012expectancy}, task facets \cite{li2008faceted}, tasks defined on intents (e.g., informational, transactional, and navigational \cite{broder2002taxonomy}; well-defined or ill-defined \cite{ingwersen2005turn}; lookup, learn, or investigate \cite{marchionini2006exploratory}), the hierarchical structure of tasks \cite{xie2008interactive}, the characteristics of tasks, and the attributes of task-searcher interaction, e.g., task difficulty and, of course, a focus in this article, task complexity \cite{kim2006task,bystrom1995task}.

\begin{figure}[t!]
  \centering
  \includegraphics[width=8cm]{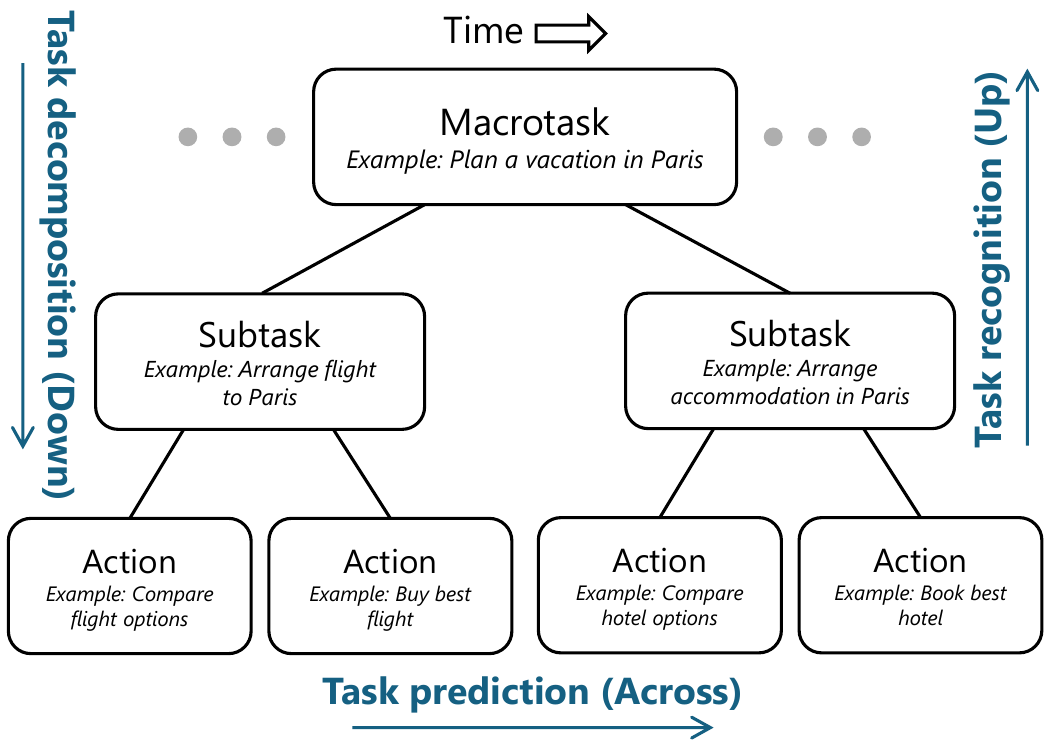}
  \caption{Task tree representation for a complex task involving planning a vacation to Paris, France. The tree depicts different task granularities (macrotask, subtask, action) and different task applications (decomposition, prediction, recognition) as moves around the tree. Time progresses from left to right via a sequence of searcher actions (queries, result clicks, pagination, etc.). Only actions are observable in traditional search engines. Aspects of subtasks and macrotasks may be observable to AI agents when searchers provide higher-level descriptions of their goals in natural language.}
  \label{task_tree}
\end{figure}

As a useful framing device to help conceptualize tasks and develop system support for them, tasks can be represented as \emph{trees} comprising macrotasks (high level goals), subtasks (specific components of those goals), and actions (specific steps taken by searchers toward the completion of those components) \cite{shah2023taking}. Figure \ref{task_tree} presents an example of a ``task tree'' for a task involving an upcoming vacation to Paris, France. Examples of macrotasks, subtasks, and actions are included. Moves around this tree correspond to different task applications such as task recognition (up), task decomposition (down), and task prediction (across). Only actions (e.g., queries, clicks, and so on) are directly observable to traditional search engines. However, with recent advances in AI agents (primarily more fully supporting natural language interactions to improve alignment between searchers and AI agents, but also a growing system awareness of short- and long-term contexts), more aspects of macrotasks and subtasks are becoming visible to search systems and more fully understood by those systems. Challenges in working with tasks include how to represent them within search systems, how to observe more task-relevant activity and content to develop richer task models, and how to develop task-oriented interfaces that place tasks and their completion at the forefront of user engagement. Task complexity deserves a special focus in this article given the challenges that searchers can still face with complex tasks and the significant potential of AI to help searchers resolve complex tasks.

\subsection{Complex Search Tasks}
Recent estimates suggest that half of all Web searches are not answered.\footnote{\url{https://blogs.microsoft.com/blog/2023/02/07/reinventing-search-with-a-new-ai-powered-microsoft-bing-and-edge-your-copilot-for-the-web/}} Many of those searches are connected to complex search tasks. These tasks are ill-defined and/or multi-step, span multiple queries, sessions, and/or devices, and require deep engagement with search engines (many queries, backtracking, branching, etc.) to complete them \cite{hassan2014supporting}. Complex tasks also often have many facets and cognitive dimensions, and are closely connected to searcher characteristics such as domain expertise and task familiarity \cite{sarkar2021integrated,white2018opportunities}.

To date, there have been significant attempts to support complex search tasks via humans (e.g., librarians, subject matter experts) and search systems (both general Web search engines and those tailored to specific industry verticals or domains). The main technological progress so far has been in areas such as query suggestion and contextual search, with new experiences also being developed that utilize multiple devices, provide cross-session support, and enable conversational search. We are now also seeing an emerging wave of search-related technologies in the area of generative AI \cite{najork2023generative}.

Before proceeding, let us dive into these different types of existing and emerging search support for complex tasks in more detail.

\begin{itemize}[leftmargin=*]
    \item \emph{Suggestions, personalization, and contextualization: } Researchers and practitioners have long developed and deployed support such as query suggestion and trail suggestion, e.g., \cite{hassan2014supporting,singla2010studying}, including providing guided tours \cite{trigg1988guided} and suggesting popular trail destinations \cite{white2007studying} as ways to find relevant resources. This coincides with work on contextual search and personalized search, e.g., \cite{bennett2012modeling,teevan2005personalizing,white2013enhancing}, where search systems can use data from the current searcher such as session activity, location, reading level, and so on, and the searcher's long-term activity history, to provide more relevant results. Search engines may also use cohort activities to help with cold-start problems for new users and augment personal profiles for more established searchers \cite{teevan2009discovering,yan2014cohort}.
    \item \emph{Multi-device, cross-device, and cross-session: } Devices have different capabilities and can be used in different settings. Multi-device experiences, e.g., \cite{white2019multi}, utilize multiple devices \emph{simultaneously} to better support complex tasks such as recipe preparation, auto repair, and home improvement that have been decomposed into steps manually or automatically \cite{zhang2021learning}. Cross-device and cross-session support \cite{agichtein2012search, wang2013characterizing} can help with ongoing/background searches for complex tasks that persist over space and time. For example, being able to predict task continuation can help with ``slow search'' applications that focus more on result quality than the near instantaneous retrieval of search results \cite{teevan2014slow}.
    \item \emph{Conversational experiences and generative AI: } Natural language is an expressive and powerful means of communicating intentions and preferences with search systems. The introduction of clarification questions on search engine result pages (SERPs) \cite{zamani2020generating}, progress on conversational search \cite{gao2023neural}, and even ``conversations'' with documents (where searchers can inquire about document content via natural language dialog) \cite{ter2020conversations}, enable these systems to engage more fully with searchers to better understand their tasks and goals. There are now many emerging opportunities to better align search systems and their users, and support more tasks, via large-scale foundation models such as OpenAI's GPT,\footnote{\url{https://openai.com/gpt-4}} Google's Gemini,\footnote{\url{https://gemini.google.com}} and Meta's Llama,\footnote{\url{https://llama.meta.com}} including offering conversational task assistance via chatbots such as ChatGPT.\footnote{\url{https://openai.com/chatgpt}}
\end{itemize}

All of these advances, and others, have paved the way for the emergence of a new class of generative-AI-powered assistive agents that can help people make progress in their complex search tasks.

\section{AI Agents}
Agents are applications of modern AI (foundation models, etc.) to help people with complex cognitive tasks. At Microsoft, we refer to these as \emph{copilots}, which work alongside humans to empower them and amplify their cognitive capabilities.\footnote{https://copilot.microsoft.com} Copilots have conversational user interfaces and their users engage with them via natural language, they are powered by foundation models such as GPT-4, they are extensible with skills/tools/plugins, and they are scoped to specialized domains or applications (including search). Copilots are designed to to keep humans at the center of the task completion process and augment human capabilities to help them complete a broader range of tasks in less time and with less effort.

\begin{figure}[t!]
  \centering
  \includegraphics[width=\linewidth]{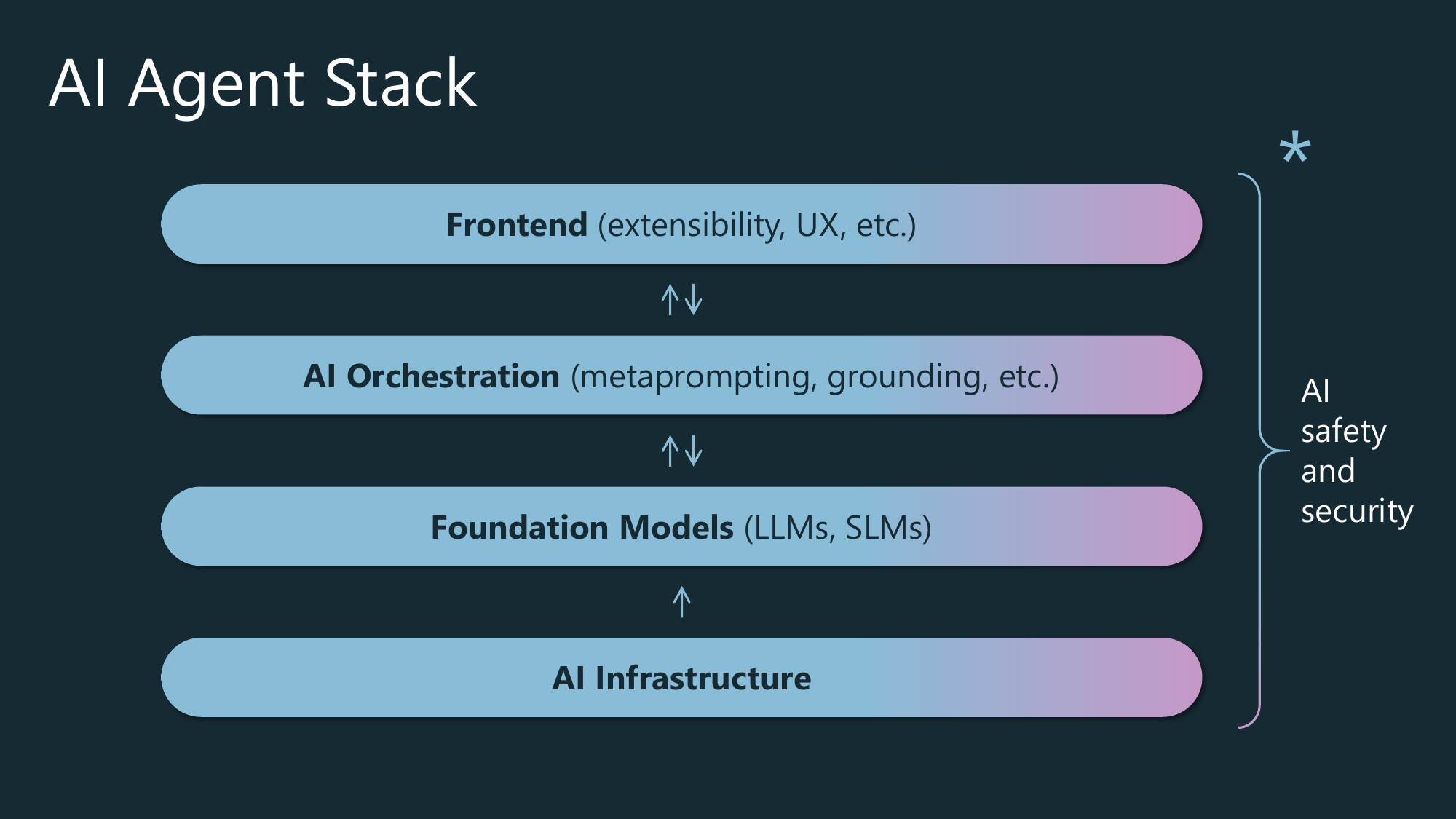}
  \caption{AI agent stack depicting the various layers and the important role of AI safety and security across the stack. Foundation models can be either large language models (LLMs), with trillions of parameters, or small language models (SLMs), with just a few billion parameters. The star (*) symbolizes that there can be multiple cooperating agents as discussed later in Section \ref{multiagent}.}
  \label{copilot_stack}
\end{figure}

The general AI agent stack (Figure \ref{copilot_stack}) contains four layers: \begin{enumerate*}
\item The \emph{frontend}, covering the user experience and extensibility with plugins, enabling developers to provide additional visible tools to the agent;
\item The \emph{AI orchestration} layer that handles the internal information flows, prompting, grounding, executing the plugins and processing their responses, among other things; 
\item Agents leverage the power of large \emph{foundation models} that can be provided to the developer as is, specialized to specific tasks, domains, or applications, or developers can bring their own models to use to power agent functionality, and;
\item This all runs on top of massive scale \emph{AI infrastructure} hosted in the cloud on platforms such as Microsoft Azure, Google Cloud, and Amazon Web Services.
\end{enumerate*}
Underpinning all of this is a need for a strong commitment to responsible AI, which ensures that agents are safe, secure, and transparent. We can do this via an iterative, layered approach with mitigations spanning the model, prompts, grounding, and the user experience.

AI agents can, among other things, help users attain goals, maximize utility, and perform automated functions. Examples of these agents include the Apple Siri and Amazon Alexa personal digital assistants that can answer questions and assist with task management, GitHub Copilot,\footnote{\url{https://github.com/features/copilot}} an AI pair programmer that has been shown to reduce developer effort, enable more task success, and significantly expedite task completion \cite{bird2022taking}, and Auto-GPT,\footnote{\url{https://auto-gpt.ai/}} a fully-autonomous agent that can decompose tasks into sub-tasks and execute them independently on a user's behalf to support goal attainment.

AI agents are also emerging in search systems. Popular Web search engines such as Bing and Google are adding agent functionalities in the form of conversational assistance; Bing has Copilot, mentioned earlier, Google has Gemini, a similar service. In search, agents can help searchers to tackle a broader range of tasks than information finding and go deeper than surface (SERP-level) interactions with content by synthesizing answers on the searcher's behalf. They also enable searchers to communicate their intents and goals more directly. Returning to the task tree  (Figure \ref{task_tree}), the focus on engaging agents via natural language interactions allows both searchers and systems to consider higher-level task representations (macrotasks, subtasks) in addition to the more granular actions (queries, result clicks, pagination, and so on) that searchers already perform when engaging with traditional search engines.

\subsection{Agents in Search}
Agents and chat experiences are a complement not a replacement to traditional search engines. Search engines have existed for decades and serve a valuable purpose: providing near instantaneous access to answers and resources for a broad range of search requests. These existing and emerging modalities can and should work well together to help searchers tackle a wider range of tasks.

\begin{figure}[t!]
  \centering
  \includegraphics[width=8cm]{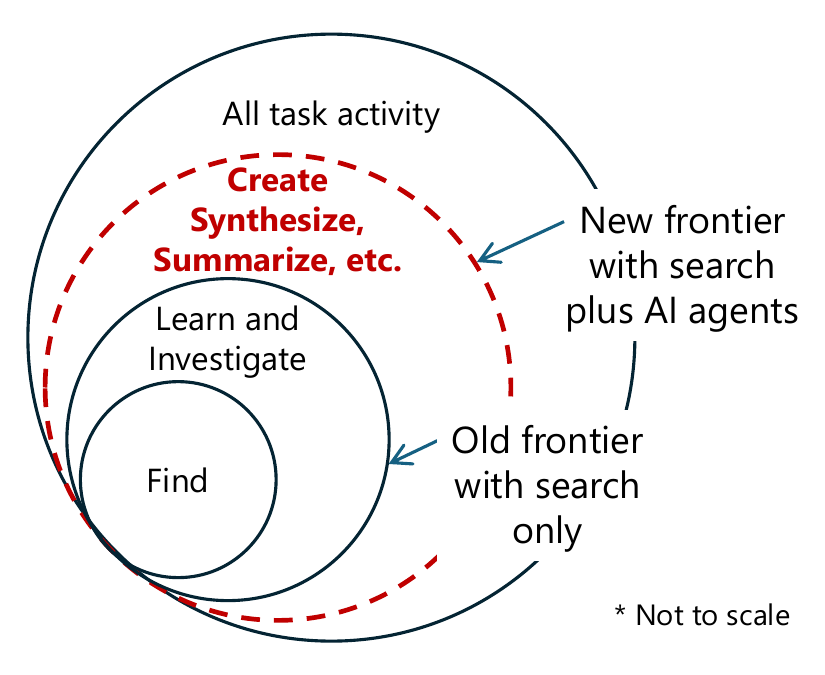}
  \caption{Advancing the search frontier. Visualizing the set of possible tasks that can be tackled with search only today (i.e., finding, learning, and investigating) plus the expansion in the frontier into support for higher-order task activities with the addition of AI agents (e.g., adding emerging AI support for creative inspiration, synthesis, and summarization).}
  \label{search_frontier}
\end{figure}

The ability of agents to better understand intentions and provide assistance beyond fact finding and basic learning/investigation will advance the \emph{search frontier} (i.e., what search systems are capable of and what types of tasks they can support), broadening the range of tasks that searchers can complete, e.g., direct support for tasks requiring creative inspiration, summarizing existing perspectives, or synthesizing those perspectives to generate new insight (Figure \ref{search_frontier}). This all moves us toward more intelligent search systems that can help with all-task completion, covering the full universe of tasks for which people might need search support, including actuation capabilities to act on tasks in the digital and physical worlds.

One way to define the range of tasks that agents can support is though Bloom's taxonomy of learning objectives \cite{krathwohl2002revision}. Creation is at the pinnacle of that taxonomy and we have only scratched the surface in creativity support with next-word prediction through transformer models \cite{devlin2018bert}. We are already seeing expansions into content types beyond text (images, video, audio, and so on) and could consider support for other creative tasks including planning, analysis, and invention. There are also many other layers in Bloom's taxonomy (e.g., evaluation - help searchers make judgments and decisions, application - help searchers complete new tasks, understanding - explain ideas and concepts to accelerate learning) that could form the basis for future search frontiers.

\begin{figure*}[h]
  \centering
  \includegraphics[width=\linewidth]{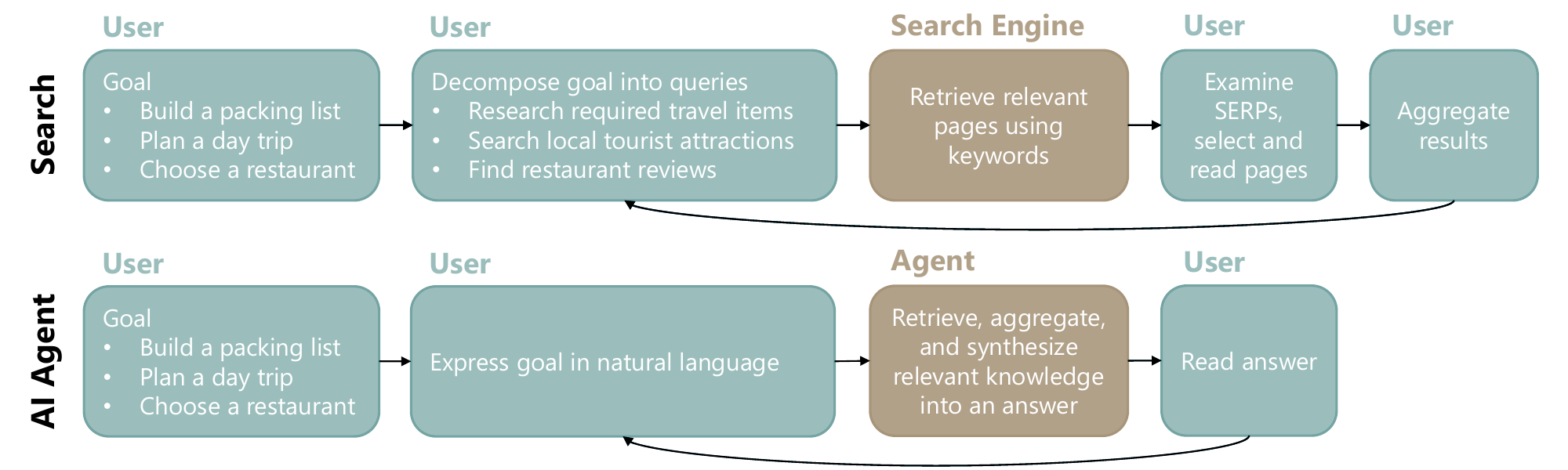}
  \caption{Information interactions in a traditional search engine versus an AI agent.}
  \label{interaction_flows}
\end{figure*}

Beyond greater capabilities, the introduction of AI agents into search will also change how people will engage with search systems. In agents, the mode of interaction is primarily natural language, with some recent expansion toward other content types for both inputs and outputs via the introduction of image and video generation models, e.g., Stable Diffusion,\footnote{https://stability.ai/stable-image} DALL$\cdot$E,\footnote{\url{https://openai.com/dall-e-3}} and Sora.\footnote{\url{https://openai.com/sora}} Agents can generate direct answers, with source attribution for provenance, to build trust with users, and to drive traffic back to content creators, which is important to incentivize further content creation that fuels future foundation models. The overall search interaction flow is also different between search engines and AI agents. When using agents, searchers do not need to decompose their goal into sub-goals or sub-queries, examine SERPs and landing pages, and aggregate/synthesize relevant knowledge from retrieved information. Continuing our running example macrotask of vacation planning from earlier, Figure \ref{interaction_flows} has a comparison of information interaction in the two modalities for some task-related goals. In AI agents, the responsibility for generating answers is delegated by the searcher to the system, which poses challenges in terms of human control and human learning, as we will discuss later in this article.

\subsection{Adding Agents to Search Engines}
It is not practical nor necessary to deploy AI agents for all search tasks. Foundation model inference is expensive at massive scale and search engine algorithms have been honed over decades to provide relevant results for a broad range of tasks (e.g., navigation, fact-finding). Conversational interfaces are less familiar for searchers and it will take time for searchers to adapt to this way of searching. Traditional search engines are sufficient when searchers know exactly what they want. Agents are helpful for more complex search tasks or in situations where searchers may be struggling to find relevant information. Task complexity can be estimated using aggregate metrics such as the amount of engagement with the search engine (e.g., number of query reformulations) for similar tasks historically. As generative AI appears in more applications and searchers better understand agent capabilities, the tasks that searchers bring to agents deployed in search settings will likely evolve and expand, and may well also increase in complexity.  

We will also see a growth in search experiences that unify traditional search and copilots. In a step towards this, search engines such as Bing and Google are already integrating dynamic answers from foundation models into their SERPs for some queries (e.g., Google's so-called ``search generative experience''\footnote{http://www. google.com/sge}). In these experiences, search results and other answers can be shown together on the SERP with answers from generative AI, allowing searchers to easily engage with them as desired, including asking follow-up questions inline. There are also ways for searchers to move between modalities based on their task and personal preferences. AI agents can also provide searchers with control over other aspects, e.g., Bing offers an ability to adjust conversation style and tone, although it is not clear that searchers are sufficiently familiar with agents at this time to use these more nuanced controls effectively.

\begin{figure}[h]
  \centering
  \includegraphics[width=\linewidth]{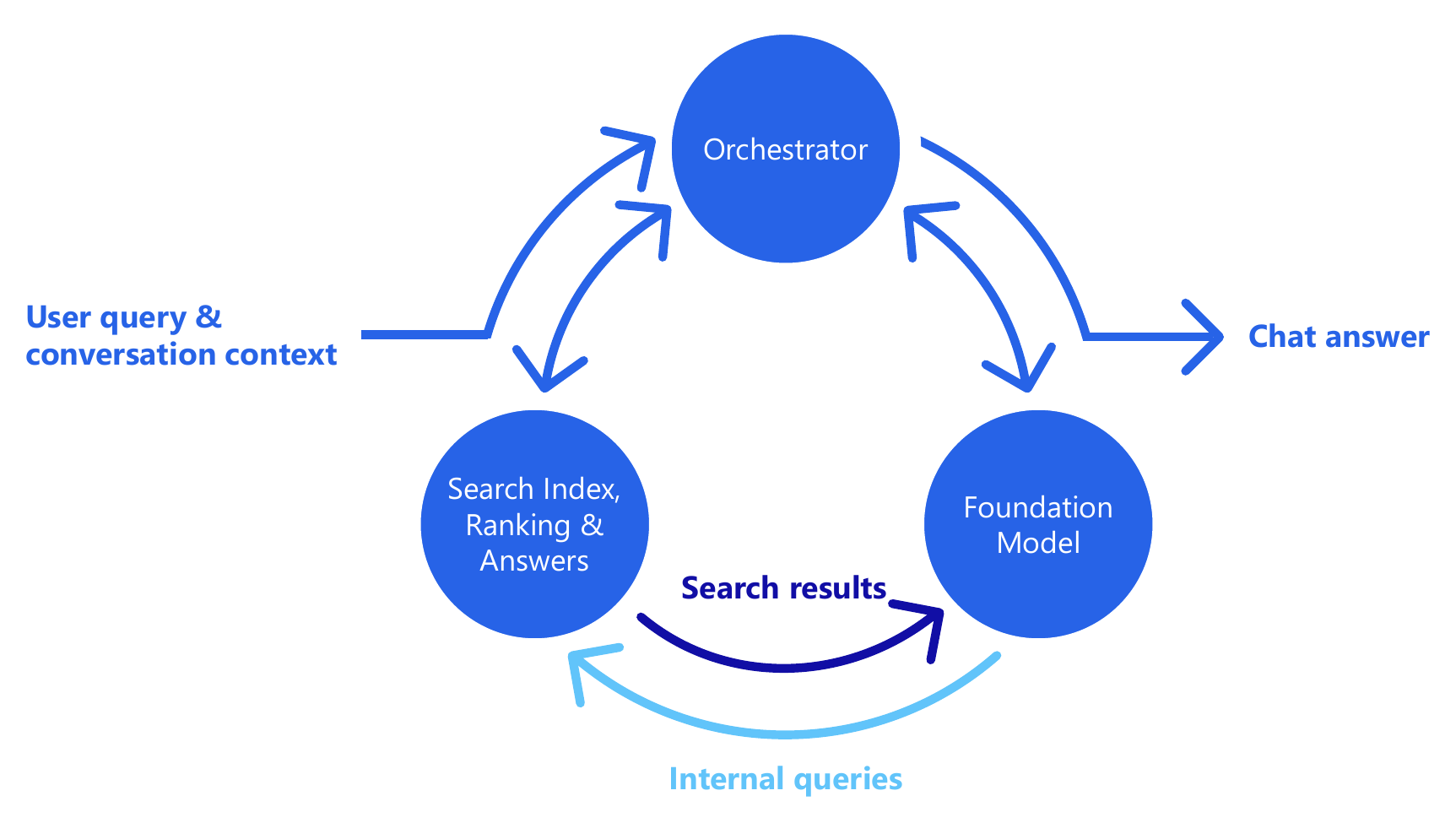}
  \caption{High-level overview of the typical generative AI search process in search engines. The query and the context are passed to the orchestrator, which coordinates with foundation model to create internal queries and generate answers. The orchestrator may also integrate content (e.g., search results, direct answers) from the search engine.}
  \label{prometheus}
\end{figure}

Search agents such as Copilot and Gemini use retrieval augmented generation (RAG) \cite{lewis2020retrieval} to ground copilot responses via timely and relevant results. This has many advantages, including:
\begin{enumerate*}
    \item There is no need to retrain massive foundation models over time;
    \item Search results provide relevant and fresh information to foundation models, and;
    \item It provides a provenance signal connecting generated content with online sources.
\end{enumerate*}
In response to a searcher prompt, the foundation model generates internal queries iteratively that are used to retrieve the results that form context for the agent answers created using generative AI. Displaying these queries to searchers inline in dialog, as in Copilot, creates greater transparency and helps build trust with searchers that the system is understanding their tasks and goals. The orchestrator can also pull in relevant instant answers from the search engine such as weather, stock, and sports, and display those in copilot responses instead of or in addition to the answers generated by the foundation model. Figure \ref{prometheus} shows the high-level search process from query (+ conversation context) to the answer, and the role of various key system components.

AI agents also enable search engines to support more complex search tasks. Using search alone would require more searcher effort to examine search results and manually generate answers or insights (see recent work on the Delphic costs and benefits of search \cite{broder2023delphic}). Of course, there are different perspectives on task complexity, e.g., the agent perspective (denoting the amount of computation, requests, etc. required for the system to complete the task) and the searcher perspective (denoting the amount of manual effort required for the human searcher to generate an answer and complete a task). Table \ref{task_complexity_tab} considers the task complexity from these two different perspectives and (again drawing from Bloom's taxonomy) provides some current, anecdotal examples of the types of tasks that both searcher and systems may find to be more or less complex. Assuming that foundational model costs will drop and sophistication will increase, we focus here on the task complexity for searchers.
    
\begin{table}
    \caption{Anecdotal examples of high-level tasks from Bloom's taxonomy \cite{krathwohl2002revision} of varying complexities from searcher/copilot perspectives. Tasks such as `Find' and `Analyze' have similar complexities for both humans and machines. It is easier for machines to create content than for humans, but more difficult for machines to verify the correctness of information.}
    \centering
    \begin{tabular}{|c|c|c|c|}
        \hline
        \multicolumn{2}{|c|}{\multirow{2}{*}{}} & \multicolumn{2}{c|}{\textbf{Searcher}}\\
        \cline{3-4}
        \multicolumn{2}{|c|}{} & \textbf{\textit{Low}} & \textbf{\textit{High}}\\
        \hline
        & \textbf{\textit{Low}} & Find & Create \\
        & & Recognize & Evaluate \\
        & & List & Compare \\
        & & Define & Predict\\
        \cline{2-4}
        \multirow{-9}{*}{\textbf{AI Agent}} & \textbf{\textit{High}} & Verify & Analyze \\
        & & Decide & Investigate \\
        & & Teach & Solve \\
        & & Plan & Invent\\
        \hline
    \end{tabular}
    \label{task_complexity_tab}
\end{table}

\section{Challenges}
Despite the promise of AI agents to dramatically improve information literacy, there are significant challenges in deploying them in search systems at scale and we must find ways to overcome these challenges. Those include issues with the agent output shown in response to searcher requests, the impacts that the agents can have on searchers, and shifts in the degree of control that humans have in the search process that come from introducing agents.

\begin{itemize}[leftmargin=*]
    \item \emph{Hallucinations: } Searchers rely a lot on the answers from agents, but those answers can be erroneous or non-sensical. So-called ``hallucination''  is a well-studied problem in foundation models \cite{ji2023survey}. Agents can hallucinate for many reasons. One of the main reasons being gaps in the training data. RAG, discussed earlier, is a way to help address this by ensuring that the agent has access to up-to-date, relevant information at inference time to help ground its responses. Injection of knowledge from other external sources, such as knowledge graphs and Wikipedia, can also help improve the accuracy of agent responses. An issue related to agents surfacing misinformation is toxicity (i.e., offensive or harmful content), which can also be present in the agent output, and must be mitigated before answers are shown to searchers.
    
    \item \emph{Biases: } Biases in the training data, e.g., social biases and stereotypes \cite{liang2021towards}, affect the output of foundation models and hence the answers provided by agents. Synthesis of content from different sources can amplify biases in this data.  As with hallucinations, this is a well studied problem \cite{bommasani2021opportunities}. Agents are also subject to biases from learning from their own or other AI-generated content (via feedback loops); biased historical sequences lead to biased downstream models. Agents may also amplify existing cognitive biases, such as confirmation bias, by favoring responses that are aligned with searchers' existing beliefs and values, and by providing responses that are optimized to keep searchers engaged with the agent, regardless of the consequences for the searcher.

    \item \emph{Human learning: } Learning may be affected/interrupted by the use of AI agents since they remove the need for searchers to engage as fully with the search system and the information retrieved. Learning is already a core part of the search process \cite{rieh2016towards,vakkari2016searching,marchionini2006exploratory}. Both exploratory search and search as learning involve considerable time and effort in finding and examining relevant content. While this could be viewed as a cost, this deep exposure to content also helps people learn. As mentioned earlier, agent users can ask richer questions (allowing them to specify their tasks and goals more fully) but they then receive synthesized answers generated by the agent, creating fewer, new, or simply different learning opportunities for humans that must be understood.

    \item \emph{Human control: } Supporting search requires considering the degree of searcher involvement in the search process, which varies depending on the search task \cite{bates1990should}. Agents enable more strategic, higher-order actions (higher up the ``task tree'' from Figure \ref{task_tree} than typical interactions with search systems). It is clear that searchers want control over the search process. They want to know what information is/not being included and why. This helps them understand and trust system output. As things stand, searchers delegate full control of answer generation to the AI, but the rest is mixed, i.e., less control of search mechanics (queries, etc.) but more control of task specifications (via natural language and dialog). There is more than just a basic tension between automation and control. In reality, it is not a zero sum game. Agent designers need to ensure human control while increasing automation \cite{shneiderman2022human}. New frameworks for multi-agent task completion are moving in this direction \cite{wu2023autogen,li2024camel}, with agents and humans working together synergistically to decompose and tackle complex tasks.
\end{itemize}

Overall, these are just a few of the challenges that affect the viability of AI agents in search settings. There are other challenges, such as searchers' deeply ingrained search habits that may be a barrier to their adoption of new search functionality, despite the clear benefits to them from embracing agent technologies.


\section{Opportunities}
\begin{figure}[t!]
  \centering
  \includegraphics[width=\linewidth]{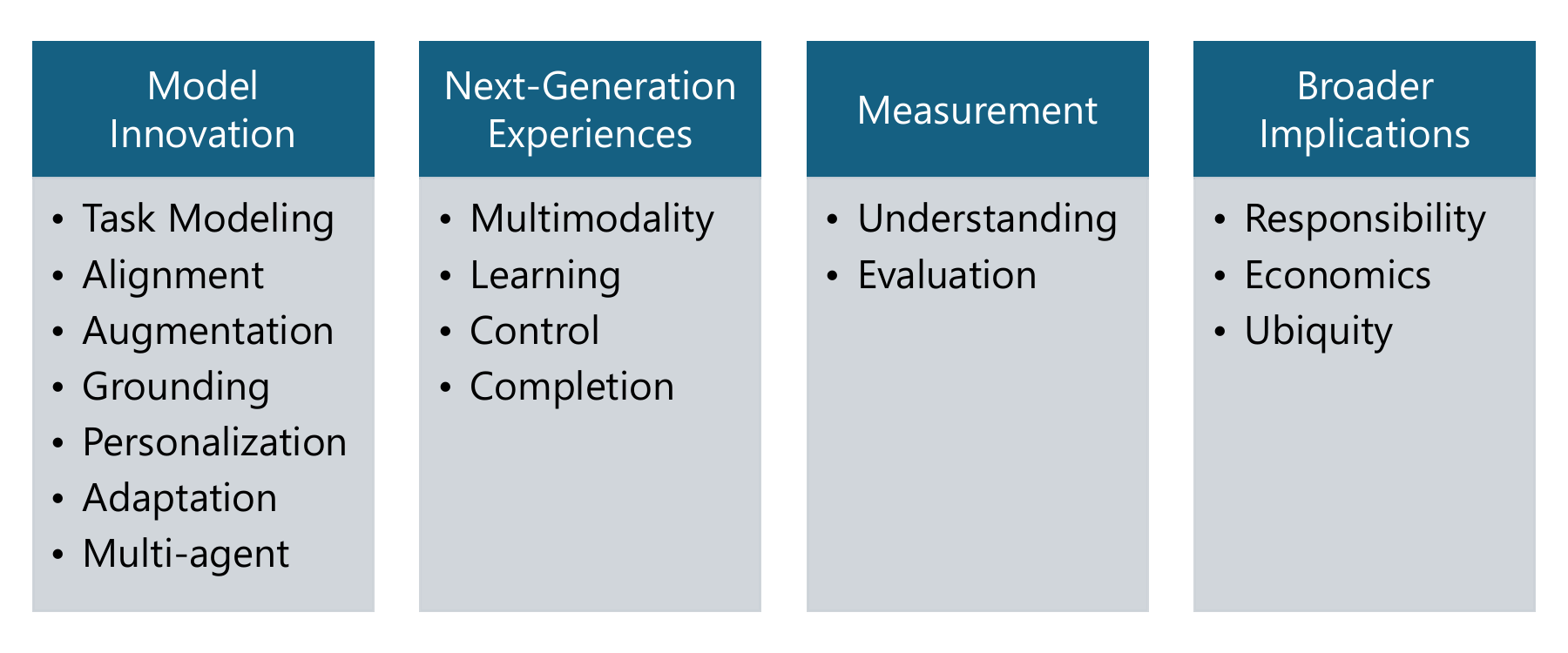}
  \caption{Selected opportunities for progress on AI agents in search settings. These opportunities are only a first step in this direction but the figure highlights the many avenues for impact in this area.}
  \label{futures}
\end{figure}

For some time, scholars have argued that the future of information interaction will involve personal search assistants with advanced capabilities, including natural language input, rich sensing, user/task/world models, and reactive and proactive experiences \cite{white2016interactions}. Technology is catching up with this vision. Opportunities going forward can be grouped into four areas: 
\begin{enumerate*}
    \item Model innovation;
    \item Next-generation experiences;
    \item Measurement, and;
    \item Broader implications.
\end{enumerate*}
The opportunities are summarized in Figure \ref{futures}. There are likely more such opportunities that are not listed here, but the long list shown in the figure is a reasonable starting point for scientists and practitioners interested in working in this area.

\subsection{Model Innovation}
There are many opportunities to better model search situations and augment and adapt foundational models to better align with searchers' tasks and goals, and provide more accurate answers. Agents can leverage these model enhancements to improve the support that they provide for complex search tasks. We now present more detail on each opportunity, including a one sentence summary (shown in italics at the beginning of each subsection).

\subsubsection{Task modeling}\label{model_innovation} \emph{Build richer task models that more fully represent tasks and task contexts.} This includes how we infer tasks and intent (e.g., from textual content of search process, from user-system interactions, from other situational and contextual information such as location, time, and application usage) and how we represent those tasks internally (e.g., as a hierarchy (Figure \ref{task_tree}) or a more abstract representation (semantic vectors, graph embeddings, Markov models, and so on)). We also need to be able to estimate key task characteristics, such as task complexity, which, in one use, can help search systems route requests to the most appropriate interaction modality. In addition, we need to find ways for agents to collect more (and more accurate) user/world knowledge, both in general and specifically related to the task at hand. A better understanding of the short- and long-term task context will help agents more accurately model the tasks themselves.

\subsubsection{Alignment} \emph{Develop methods to continuously align agents to tasks/goals/values via feedback}, e.g., conversation content as feedback (e.g., expressed positive sentiment or searchers expressing gratitude to the agent) or explicit feedback on agent answers via likes and dislikes. The performance of agents that are missing alignment will remain fixed over time. Agents need application-aligned feedback loops to better understand searcher goals and tasks and use that feedback to continuously improve answer accuracy and relevance. Beyond research on fine-tuning foundation models from human feedback (e.g., likes/dislikes) \cite{ziegler2019fine}, we can also build on learnings from research on implicit feedback in IR, including work on improving ranking algorithms via SERP clicks \cite{joachims2002optimizing} and developing specialized interfaces to capture user feedback \cite{white2005study}.
                
\subsubsection{Augmentation} \emph{Enhance agents with relevant external knowledge and enhanced tools and capabilities.} As mentioned earlier, RAG is a common form of knowledge injection for foundation models. Relevance models are tuned to maximize user benefit, not for agent consumption. We need to evaluate whether this difference is meaningful practically and if so, develop new ranking criteria that consider the intended consumer of the search results (human or machine). Despite their incredible capabilities, foundation models still have shortcomings that manifest in the agents that use them. We need to understand these shortcomings through principled evaluation and find ways to leverage external skills/plugins to address them. Agents must find and recommend skills per task demands \cite{white2018skill}, e.g., invoking Wolfram for computational assistance. We can also integrate tool use directly into tool-augmented models, e.g., \emph{Toolformer} \cite{schick2023toolformer}, that can teach themselves to use tools. Models of task context may also be incomplete and we should invest in ways to better \emph{ground} agent responses via context, e.g., richer sensing, context filtering, and dynamic prompting.

\subsubsection{Grounding}\label{grounding} \emph{Apply use-case specific information to reduce hallucinations, build trust, and support content creators}. It is in the interests of agents, searchers, and content creators (and providers and advertisers) to consider the source of the data used in generating answers. Provenance is critical and agents should provide links back to relevant sources (preferably with specific details/URLs not generalities/domains) to help establish and maintain user trust, provide attribution for content creators, and drive engagement for content providers and advertisers. It also important for building trust and for supporting learning for agents to practice faithful reasoning \cite{creswell2022faithful}, and provide intepretable reasoning traces (e.g., explanations with chain-of-thought) associated with their answers. We should also consider how we integrate search within existing experiences (e.g., in other agents) to ground answers in their context of use and in more places that people seek those answers.

\subsubsection{Personalization} \emph{Develop personal agents that can understand searchers and their tasks, using personal data, privately and securely.} Searchers bring their personal tasks to search systems and agents will be no different. Here are some example personal prompts that describe the types of personal tasks that searchers might expect an agent to handle:
\begin{enumerate*}[leftmargin=*]
    \item Write an e-mail to my client in my personal style with a description of the quote in the attached doc.
    \item Tell me what’s important for me to know about the company town hall that I missed?
    \item Where should I go for lunch today?
\end{enumerate*}
These tasks span creation, summarization, and recommendation and quickly illustrate the wide range of expectations that people may have from their \emph{personal} agents. As part of developing such personalized AI support, we need to: 
\begin{enumerate*}
    \item Study foundation model \emph{capabilities}, including their ability to identify task-relevant information in personal data and activity histories, and model user knowledge in the current task and topic, and
    \item Develop core \emph{technologies}, including infinite memory, using relevant long-term activity (in IR, there has been considerable research on relevant areas such as re-finding \cite{tyler2010large} and personalization \cite{teevan2005personalizing}); context compression, to fit more context into finite token limits (e.g., using turn-by-turn summarization rather than raw conversational content); privacy, including mitigations such as differential privacy and federated learning, and research on machine unlearning \cite{bourtoule2021machine} to intentionally forget irrelevant information over time, including sensitive information that the searcher may have explicitly asked to be removed from trained models, and also remove irrelevant or unwanted data from agent memory.
\end{enumerate*}

\subsubsection{Adaptation}\label{adaptation} Two main forms of adaptation that we consider here are model specialization and so-called adaptive computation.

\begin{itemize}[leftmargin=*]
    \item \emph{Model specialization.} \emph{Develop specialized foundation models for search tasks that are controllable and efficient.} Large foundation models are generalists and have a wide capability surface. Specializing these models for specific tasks and applications discards useless knowledge, making the models more accurate and efficient for the task at hand. Recent advances in this area have yielded strong performance, e.g., the Orca-13B model \cite{mukherjee2023orca} uses explanation-based tuning (where the model explains the steps used to achieve its output and those explanations are used to train a small language model (SLM)) to outperform state-of-the-art models of a similar size such as Vicuna-13B \cite{chiang2023vicuna}. Future work could explore guiding specialization via search data, including anonymized large-scale search logs, and as well as algorithmic advances in searcher preference modeling and continual learning.
    \item \emph{Adaptive computation.} \emph{Develop methods to adaptively apply different models per task and application demands.} Adaptive compute involves using multiple foundation models (e.g., GPT and a specialized model) each with different inference-time constraints, primarily around speed, capabilities, and cost, and learning which model to apply for a given task. The specialized model can backoff to one or more larger models as needed per task demands. The input can be the task plus the constraints of the application scenario under which the model must operate. Human feedback can also be used to refine the adaptation strategy over time \cite{zhang2023ecoassistant}. 
\end{itemize}
These adaptation methods will yield more effective and more efficient AI capabilities that agents can use to help searchers across a range of settings, including in offline settings (e.g., on-device only).

\subsubsection{Multi-Agent} \label{multiagent} \emph{Utilize multiple specialized agents working together and with humans to help complete a search task.} Multiple agents have been shown to help encourage divergent thinking \cite{liang2023encouraging}, improve factuality and reasoning \cite{du2023improving}, and provide guardrails for AI systems \cite{wu2023empirical}. Multi-agent systems such as \emph{AutoGen} \cite{wu2023autogen} and \emph{CAMEL} \cite{li2024camel} orchestrate communication between agents to help users complete tasks more effectively. These systems could be used in search settings to, for example, retrieve relevant resources from diverse sources, improve answer correctness by critiquing and refining AI-generated output, improve human decision making by presenting alternative solutions, and even automate the completion of some tasks or sub-tasks, with humans in the loop throughout.

\subsection{Next-Generation Experiences}
Advancing models is necessary but not sufficient given the central role that interaction plays in the search process \cite{white2016interactions}. There are many opportunities to develop new search experiences that capitalize on agent capabilities while keeping searchers in control.

\subsubsection{Multimodality} \emph{Develop experiences bridging (at least) the search and agent (chat) modalities, offering explanations and suggestions.} Given how entrenched and popular traditional search is, it is likely that some form of query-result interaction will remain a core part of how we find information online. Future, agent-enhanced experiences may reflect a more seamless combination of the two interaction modalities in a unified experience. Both Google and Bing are taking a step in that direction by unifying search results and agent answers in a single interface. Explanations on which modality and style (e.g., creative, balanced, and precise) perform best and when will help searchers make decisions about which modalities and settings to use. Modality recommendation given task is also worth exploring: simple tasks may only need traditional search, whereas complex tasks may need agents. Contextualization and personalization will also play an important part in deciding how much information is needed from the searcher (incurring interaction cost but yielding greater control) and how much can be reliably inferred from signals already available to the system. Related to this are opportunities around conversation style suggestion given the current task, e.g., fact-finding task or short reply (needs precision, when generative-AI-powered agents can often be verbose) and generating new content (needs creativity, when agent responses can often be unoriginal or bland). Search providers could also consider offering a single point of entry and an automatic routing mechanism to direct requests to the correct modality given inferences about the underlying task (e.g., from Section \ref{model_innovation}) and the appropriateness of each of the modalities for that task. Beyond search and chat, other modalities to help support complex search tasks may include third-party tools and applications, bespoke user interfaces (e.g., tailored dynamically by the agent to the task at hand), interactive visualizations, and proactive recommendations.

\subsubsection{Human Learning} \emph{Develop agents that can detect learning-related search tasks and support relevant learning activities.} As mentioned earlier, agents can remove or change human learning opportunities by their automated generation and provision of answers. Learning is a core outcome of information seeking \cite{vakkari2016searching,marchionini2006exploratory,dervin1998sense}. We need to develop agents that can detect learning and sensemaking tasks, and support relevant learning activities via agent experiences that, for example, provide detailed explanations and reasoning, offer links to learning resources (e.g., instructional videos), enable deep engagement with task content (e.g., via relevant sources), and support specifying and attaining learning objectives. 

\subsubsection{Human Control} \emph{Better understand control and develop agents with control while growing automation.} Control is an essential aspect of searcher interaction with agents. Agents should consult humans to resolve or codify value tensions. Agents should be in collaboration mode by default and must only take control with the permission of stakeholders. Experiences that provide searchers with more agency are critical, e.g., adjust specificity/diversity in agent answers, leading to less generality and less repetition. As mentioned in Section \ref{grounding}, citations in answers are important. Humans need to be able to verify citation correctness in a lightweight way, ideally without leaving the user experience; Gemini now offers an ability to manually dive deeper and verify answers. We also need a set of user studies to understand the implications of less control of some aspects (e.g., answer generation), more control over other aspects (e.g., macrotask specification), and control over new aspects, such as conversation style and tone, as with Copilot.

\subsubsection{Completion} \emph{Agents should help searchers complete tasks while keeping searchers in control.} We need to both expand the search frontier by adding/discovering more capabilities of foundation models that can be surfaced through agents \emph{and} deepen task capabilities so that agents can help searchers better complete more tasks. We can view skills and plugins as actuators of the digital world and we should help foundation models fully utilize them. We need to start simple (e.g., reservations), learn and iterate, and increase task complexity as model capabilities improve with time.

The standard mode of engagement with AI agents is reactive; users send requests and the agents respond. Agents should ideally have a \emph{dynamic interaction model} that tailors the interface to the task and the context. They can take initiative, with permission, and provide updates for standing tasks, or they could also offer proactive suggestions or take actions directly when agent uncertainty is low. Agents can also help to support task planning (decomposition, prioritization, and so on) for complex tasks such as travel or events. AI can already help complete repetitive tasks, e.g., action transformers, trained on digital tools\footnote{\url{https://www.adept.ai/blog/act-1}} and create and apply ``tasklets'' (user interface scripts) learned from websites \cite{li2021glider}.

Given the centrality of information interaction in search task completion, it is important to focus sufficient attention on interaction models and experiences in AI agents. In doing so, we must also carefully consider the implications of critical decisions on issues that affect AI in general such as control and automation.

\subsection{Measurement}
Another important direction is in measuring AI agent performance, understanding agent impact and capabilities, and tracking agent evolution over time. Many of the challenges and opportunities in this area also affect the evaluation of foundation models in general (e.g., non-determinism, saturated benchmarks, inadequate metrics).

\subsubsection{Understanding} \emph{Deeply understand agent capabilities and agent impact on searchers and on their tasks.} We have only scratched the surface in understanding AI agents and their impact. A deeper understanding takes a few forms, including:
\begin{enumerate*}
    \item \emph{User understanding}: Covering mental models of agents and effects of bias (e.g., functional fixedness \cite{duncker1945problem}) on how agents are adopted and used in search settings. It also covers changes in search behavior and information seeking strategies, including measuring changes in effects across modalities, e.g., search \emph{versus} agents and search \emph{plus} agents. There are also opportunities in using foundation models to understand search interactions via user studies \cite{capra2023does} and use foundation models to generate intent taxonomies and classify intents from log data \cite{shah2023using};
    \item \emph{Task understanding}: Covering the intents and tasks that agents are used for and most effective for, and;
    \item \emph{Agent understanding}: Covering the capabilities and limitations of agents, e.g., similar to the ``Sparks of AGI'' paper on GPT-4 \cite{bubeck2023sparks}, which examined foundation model capabilities in depth.
\end{enumerate*}

\subsubsection{Evaluation} \emph{Identify and develop metrics for agent evaluation, while considering important factors, and find applications of agent components for IR evaluation.} There are many options for AI agent metrics, including feedback, engagement, precision-recall, generation quality, answer accuracy, and so on. Given the task focus, metrics should likely target the task holistically (e.g., success, effort, satisfaction). In evaluating agents in search settings, it is also important to consider:
\begin{enumerate*}
    \item Repeatability: Non-determinism can make agents difficult to evaluate/debug;
    \item Interplay between search and agents (switching, joint task success, and so on);
    \item Longer term effects on user capabilities and productivity;
    \item Task characteristics: Complexity, and so on, and;
    \item New benchmarks: Agents are affected by external data, grounding, queries, and so on.
\end{enumerate*}
There are also opportunities to consider applications of agent components for IR evaluation. Foundation models can predict searcher preferences \cite{thomas2023large} and assist with relevance judgments \cite{faggioli2023perspectives}, including generating explanations for judges. Also, foundation models can create powerful searcher simulations that can mimic human behavior and values, expanding on early work on searcher simulations in IR \cite{white2005evaluating}.

Measuring agent performance is essential in understanding their utility and improving their performance over time. Agents do not function in a vacuum and we must consider the broader implications of their deployment for complex tasks in search settings.

\subsection{Broader Implications}
AI agents must operate in a complex and dynamic world. There are several opportunities beyond advances in technology and in deepening our understanding of agent performance and capabilities.

\subsubsection{Responsibility} \emph{Understand factors affecting reliability, safety, fairness, and inclusion in agent usage in search.} The broad reach of search engines means that AI agents have a critical obligation to act responsibly. Research is needed on ways to improve answer accuracy via better grounding in more reliable data sources, develop guardrails, understand biases in foundation models, prompts, and the data used for grounding, and understand how well agents work in different contexts, with different tasks, and with different people and cohorts. Red teaming, user testing, and feedback loops are all needed to determine emerging risks in agents and the foundation models that underlie them. This also builds on existing work on responsible AI and responsible IR and FACTS-IR, which has studied biases and harms, and ways to mitigate them \cite{olteanu2021facts}.

\subsubsection{Economics} \emph{Understand and expand the economic impact of agents in search.} This includes exploring new business models which agents will create beyond information finding. Advancing the search frontier from information finding deeper into task completion (e.g., into creation and analysis) creates new business opportunity. It also unlocks new opportunities for advertising, including advertisements that are shown inline with dialog/answers and contextually relevant to the current conversation. There is also a need to more deeply understand the impact of agents on content creation and search engine optimization. Content attribution is vital in such scenarios to ensure that content creators (and advertisers and publishers) can still generate returns. We should avoid the so-called ``paradox of reuse'' \cite{vincent2022paradox} where lower visits to online content leads to less content being created which in turn leads to worse models over time. Another important aspect of economics is the cost-benefit trade-off and is related to work on adaptation (Section \ref{adaptation}). Large model inference is expensive and unnecessary for many applications. This cost will reduce with optimization, for which model specialization and adaptive computation can help, as does the emergence of high-performing SLMs of a few billion parameters, such as Phi, trained on highly-curated data \cite{gunasekar2023textbooks}.

\subsubsection{Ubiquity} \emph{Agent integrations to model and support complex search tasks.} AI agents must co-exist with the other parts of the application ecosystem. Search agents can be integrated into applications such as Web browsers (offering in-browser chat, editing assistance, summarization) and productivity applications (offering support in creating documents, emails, presentations, and so on). These agents can capitalize on application context to do a better job of answering searcher requests. Agents can also span applications through integration with the operating system. This enables richer task modeling and complex task support, since such tasks often involve multiple applications. Critically, we must do this privately and securely to mitigate risks for searchers and earn their trust.

\subsection{Summary}
The directions highlighted in this section are just examples of the opportunities afforded by the emergence of generative AI and agents in search settings. There are other areas for search providers to consider too, such as multilingual agent experiences (i.e., foundation models are powerful and could help with language translation \cite{zhu2023multilingual,mayfield2023synthetic}), agent efficiency (i.e., large model inference is expensive and not sustainable at massive scale, so creative solutions are needed \cite{zhang2023ecoassistant}), reducing the carbon impact from running foundation models at search engine scale to serve billions of answers in AI agents \cite{everman2023evaluating}, making agents private by design \cite{yu2021differentially}, and government directives (e.g., the 2023 executive order from U.S. President Biden on AI safety and security\footnote{\url{https://www.whitehouse.gov/briefing-room/statements-releases/2023/10/30/fact-sheet-president-biden-issues-executive-order-on-safe-secure-and-trustworthy-artificial-intelligence/}}) and legislation, among many other opportunities.

\section{The Undiscovered Country}
AI agents will transform how we search. Tasks are central to people's lives and more support is needed for complex tasks in search settings. Some limited support for these tasks already exists in search engines, but agents will advance the search frontier to make more tasks actionable and help to make inroads in the ``last mile'' in search interaction: task completion \cite{white2018opportunities}. Moving forward, search providers should invest in ``better together'' experiences that utilize agents plus traditional search (plus more modalities going forward), make these joint experiences more seamless for searchers, and add more support for their use in practice, e.g., help people to quickly understand agent capabilities and potential and/or recommend the best modality for the current task or task stage. This also includes experiences where both modalities are offered separately and can be selected by searchers and those where there is unification and the selection happens automatically based on the task and the conversation context.

The foundation models that power AI agents have other search-related applications, e.g., for generating and applying intent taxonomies \cite{shah2023using} or for evaluation \cite{faggioli2023perspectives}. We must retain a continued focus on human-AI cooperation, where searchers stay in control while the degree of system support increases as needed \cite{shneiderman2022human}, and on AI safety and security. Searchers need to be able to trust agents but also be able to verify their answers with minimal effort.

Overall, the future is bright for IR, and AI research in general, with the advent of generative AI and the agents that build upon it. Agents will help augment, empower, and inspire searchers on their task journeys. Computer science researchers and practitioners should embrace this new era of assistive AI agents and engage across the full spectrum of exciting practical and scientific opportunities, both within search as we focused on in this article, and onwards into other important domains such as personal productivity \cite{bird2022taking} and scientific discovery \cite{hope2023computational}.

\balance
\bibliographystyle{ACM-Reference-Format}
\bibliography{sample-base}

\end{document}